\documentclass[]{aa}
\usepackage{graphicx}
\usepackage{txfonts}
\begin{document}
\title{
VLTI/VINCI observations of the nucleus of NGC~1068 using the 
adaptive optics system MACAO\thanks{Based on public commissioning
data released from the VLTI 
(www.eso.org/projects/vlti/instru/vinci/vinci\_data\_sets.html).
}}
\author{
M.~Wittkowski\inst{1} \and
P.~Kervella\inst{1,2} \and
R.~Arsenault\inst{1} \and
F.~Paresce\inst{1} \and
T.~Beckert\inst{3} \and
G.~Weigelt\inst{3}
}
\offprints{M.~Wittkowski, \email{mwittkow@eso.org}}
\institute{
European Southern Observatory, Karl Schwarzschild-Str. 2, 
D-85748 Garching, Germany
\and
LESIA, Observatoire de Paris, F-92195 Meudon Cedex, France
\and
Max-Planck-Institut f\"ur Radioastronomie, Auf dem H\"ugel 69, 
D-53121 Bonn, Germany}
\titlerunning{VLTI/VINCI observations of NGC~1068}
%
%\mail{M.~Wittkowski, \email{mwittkow@eso.org}}
%
\date{Received \dots; accepted \dots}
\abstract{
We present the first near-infrared $K$-band long-baseline interferometric 
measurement of the nucleus of the prototype Seyfert\,2 Galaxy NGC~1068 
with resolution 
$\lambda/B \sim$\,10\,mas 
obtained with the Very Large Telescope Interferometer (VLTI) and
the two 8.2\,m diameter Unit Telescopes UT\,2 and UT\,3.
The adaptive optics system 
MACAO (Multi Application Curvature Adaptive Optics) was employed to 
deliver wavefront-corrected beams to the $K$-band commissioning 
instrument VINCI. 
A squared visibility amplitude of 16.3\,$\pm$\,4.3\,\% was measured
for NGC~1068 at a sky-projected baseline length of 45.8\,m and 
azimuth angle 44.9\,deg. This value corresponds to a FWHM of the 
$K$-band intensity distribution of 
5.0\,$\pm$\,0.5\,mas (0.4\,$\pm$\,0.04\,pc at the distance of NGC~1068) 
if it consists of a single Gaussian component. 
Taking into account $K$-band speckle interferometry observations
(Wittkowski et al. \cite{wittkowski}; 
Weinberger et al. \cite{weinberger}; Weigelt et al. \cite{weigelt}), 
we favor a multi-component model 
for the intensity distribution where a part of the flux 
originates from scales clearly smaller 
than $\sim$\,5\,mas ($\lesssim$\,0.4\,pc), 
and another part of the flux from larger scales. 
The $K$-band emission from the small ($\lesssim$\,5\,mas) scales 
might arise from substructure of the dusty nuclear torus, 
or directly from the central accretion flow viewed through 
only moderate extinction. 
\keywords{Techniques: interferometric -- galaxies: nuclei -- 
galaxies: Seyfert -- galaxies: individual: NGC~1068}
}

\maketitle

\section{Introduction}
\begin{table*}
\caption{Calibration sequence of NGC~1068. JD is the Julian date of the 
observation, $N$ the number of processed interferograms, $B$ and $Az$ the 
projected baseline length and azimuth angle (east of north), 
and $\mu^2$ the obtained coherence factors with 
their statistical errors. The interferometric efficiency (IE) given 
in italic characters is the value adopted for the calibration 
of the NGC~1068 data.}
\label{Cal_ngc1068}
\begin{tabular}{ccccclll}
\hline
JD & $N$ & B\,(m) & $Az$ & $\mu^{2} \pm$ stat. & IE $\pm$ stat. $\pm$ syst. 
& $V^2 \pm$ stat. $\pm$ syst. & Target\\ 
\hline
2452948.717 & 250 & 45.79 & 44.87 & $0.054 \pm 0.017$ & 
$\it 0.340 \pm 0.009 \pm 0.001$ & $0.158 \pm 0.049 \pm 0.001$ & 
\object{NGC~1068} \\
2452948.722 & 214 & 46.00 & 45.13 & $0.062 \pm 0.031$ & 
$\it 0.340 \pm 0.009 \pm 0.001$ & $0.182 \pm 0.091 \pm 0.001$ & 
\object{NGC~1068} \\\hline
2452948.761 & 407 & 46.62 & 45.40 & $0.319 \pm 0.005$ & 
$0.348 \pm 0.006 \pm 0.001$ & & \object{HD\,20356}\\
2452948.767 & 438 & 46.64 & 45.62 & $0.304 \pm 0.005$ & 
$0.331 \pm 0.005 \pm 0.001$ & & \object{HD\,20356}\\
2452948.772 & 416 & 46.63 & 45.80 & $0.315 \pm 0.006$ & 
$0.344 \pm 0.007 \pm 0.001$ & & \object{HD\,20356}\\
\hline
\end{tabular}
\end{table*}
The Seyfert galaxy NGC~1068 harbors one of the brightest and closest
active galactic nuclei (AGN). Active galaxies appear as types
1 and 2, where the spectra of the former exhibit 
broad and narrow emission lines, and those of the latter 
show only narrow lines.
Antonucci \& Miller (\cite{antonuccimiller}) 
suggested that the broad-line emission regions are located
inside an optically and geometrically thick disk and that central 
continuum and broad-line photons are scattered into the line-of-sight
by free electrons above and below the disk. 
Depending on the observer's viewing angle, the broad-line emission
region is either obscured or not. 
This suggestion is now widely accepted and has evolved to the so-called 
``unified scheme of AGN'' (e.g., Antonucci \cite{antonucci}). 

Various theoretical models of the postulated dusty tori 
have been presented by, for instance, 
Krolik \& Begelmann (\cite{krolikbegelmann}),
Pier \& Krolik (\cite{pierkrolik1}), 
 Granato \& Danese (\cite{granato}), 
Efstathiou \& Rowan Robinson (\cite{efstathiou1}), 
Manske et al. (\cite{manske}), Nenkova et al. (\cite{nenkova}), 
and Vollmer et al. (\cite{vollmer}). Owing to the lack of
spatially resolved observations, the models were 
usually compared to the integrated spectrum of the core of NGC~1068.  
It turned out that torus models
with a wide variety of geometries, spatial extensions, and optical
depths are consistent with this spectrum. In addition, the nature of the
central emission source itself could not be well constrained because the 
flux spectrum of the very inner nuclear engine could not be separated 
from the emission of surrounding material.

Several high-resolution infrared observations of NGC~1068 showing a
compact central IR core and surrounding structure 
were carried out by, for instance,
Thatte et al. (\cite{thatte}), 
Rouan et al. (\cite{rouan}),
Wittkowski et al. (\cite{wittkowski}), 
Weinberger et al. (\cite{weinberger}), and 
Bock et al. (\cite{bock}).
Wittkowski et al. (\cite{wittkowski}) presented 
both a $K$-band visibility function up to spatial frequencies
corresponding to a baseline of 6\,m, as well as the first 76\,mas
resolution $K$-band image of the nucleus of NGC~1068 
obtained by bispectrum speckle interferometry.
The compact central IR core 
was resolved with a Gaussian FWHM of $\sim$\,30\,mas (2\,pc). 
New $K$-band and $H$-band bispectrum speckle interferometry
observations were performed by Weigelt et al. (\cite{weigelt}).

Very recently, the first
interferometers consisting of 8--10\,m class telescopes started
operations, and they have already succeeded in observing AGN with much
higher spatial resolutions.
The potential of optical/infrared interferometry to investigate 
AGN was recently discussed by Wittkowski et al. (\cite{wittkowski2}).
Swain et al. (\cite{swain}) reported the first $K$-band interferometric
observations of the Seyfert\,1 galaxy NGC~4151 obtained 
with the 85\,m baseline of the Keck Interferometer. 
Jaffe et al. (\cite{jaffe}) reported on the first mid-infrared
interferometric observation of the dusty torus of NGC~1068 using VLTI/MIDI. 

In the present letter, we report on the first $K$-band long-baseline 
interferometric observation of NGC~1068. 
\section{Observations and data reduction}
The NGC~1068 interferometric data were obtained with the Very Large
Telescope Interferometer (VLTI) and the $K$-band commissioning 
instrument VINCI (Kervella et al.~\cite{kervella00}; 
Kervella et al.~\cite{kervella03a}), used with the fiber-based
beam combiner MONA, on Nov. 4, 2003. The UT2-UT3 baseline
with 47\,m ground length was used. Both telescopes were equipped with 
the Multi Application Curvature Adaptive Optics (MACAO) system. 
These data were taken in the framework of the commissioning of MACAO-VLTI.

\paragraph{MACAO systems.}
For a detailed description of the MACAO systems see
Arsenault et al. (\cite{arsenault}).
The MACAO systems are four curvature adaptive optics systems 
installed at the Coud\'e focus of each UT (the systems for UTs 1 
and 4 will be installed in the fall of 2004 and 
first quarter of 2005). The Coud\'e train mirror M\,8 is replaced 
by a 60 actuator bimorph deformable mirror in a tip-tilt mount. 
The mirror M\,9 is a dichroic which transmits visible light to 
the wavefront sensor
and reflects the wavelength range from 1 to 13\,$\mu$m 
to the VLTI recombination laboratory. 
During commissioning, the two MACAO systems on UTs 2 and 3
showed very similar and consistent results. 
Strehl ratios of 65\% were obtained on bright 
guide stars ($V<11$) under optical seeing conditions of 
up to 0.8\arcsec. 
On fainter sources with $V$ magnitudes of 14.5 and 16,
Strehl ratios of 45\% (seeing 0.60\arcsec) and 25\% (seeing 0.55\arcsec)
were obtained, respectively.
For use with the single-mode fiber instrument VINCI,
a high Strehl ratio ensures a high concentration of light 
onto the 56\,mas diameter fiber core of VINCI, hence allowing the
interferometric observation of faint sources such as NGC~1068.
For both MACAO-VLTI systems the reference source was 
the nucleus of  NGC~1068 itself, and the same control parameters 
were used for both AO systems. No neutral density filters 
were used, the main loop gain was 0.50.
 The optical seeing at the time of observation was $\sim$\,0.9\arcsec.
It is difficult to estimate the MACAO performance for our NGC~1068
observations since the source is extended and an acquisition
image could not be taken.
\paragraph{Interferometric data.}
1\,000 interferograms of the nucleus of NGC~1068 were obtained
in two series of 500 scans.
\begin{figure}
\centering
\resizebox{0.98\hsize}{!}{\includegraphics[bb=0 0 360 288]{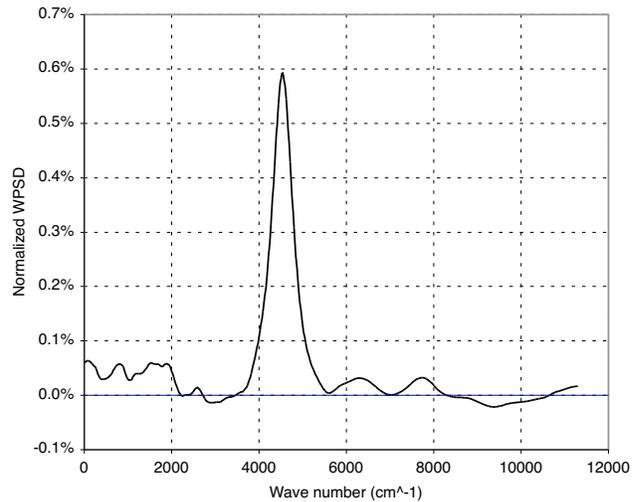}}
\caption{Average wavelets power spectral density (WPSD) of the 464
processed interferograms of NGC~1068.
The subtraction of the background noise from the processed fringes' power 
peak left no residual bias on the final WPSD. The power integration is done 
between wave numbers 2000 and 8000~cm$^{-1}$.}
\label{wl_psd_peak}
\end{figure}
The fringe frequency was 216\,Hz, a compromise 
between faint object sensitivity and immunity to the 
atmospheric piston effect.
Of the recorded interferograms, 464 were processed 
successfully by the VINCI data reduction software as described
by Kervella et al. (\cite{kervella03b}).
Table~\ref{Cal_ngc1068} lists the observational details for the NGC~1068
and the calibrator data, as well as the resulting visibility values.
Figure~\ref{wl_psd_peak} shows the background- and noise-corrected
average wavelet power spectral density of the NGC~1068 interferograms. 
The fringes' power peak around wave number 4500\,cm$^{-1}$ is not affected 
by any significant power spectrum bias, despite the faintness of the source. 
The effective wavelength of the NGC\,1068 observations is approximately
$2.18\,\mu$m.
The K\,5 giant HD\,20356 
($\theta_{\rm UD} = 1.81 \pm 0.02$\,mas) from Cohen et al.~(\cite{cohen99})
and Bord\'e et al.~(\cite{borde02})
was used as calibration star (effective wavelength $\lambda = 2.181\,\mu$m).
Observational parameters and data reduction of the calibration star
were identical to those of NGC~1068.
The systematic error induced by the calibrator on the final visibility 
values is negligible compared to the statistical error.
A calibrated squared visibility of 
$V^2 = 16.3 \pm 4.3$\,\% for NGC~1068 was obtained
for a sky-projected baseline with length $B = 45.839$\,m and 
azimuth angle $Az = 44.93\,\deg$ (east of north).
\paragraph{Photometric estimate. \label{magnitude_sect}}
Any photometric estimates using a single-mode fiber instrument
is, in general, difficult due to the large and rapid fluctuations 
of the coupling of the object light into the fiber core. However, the
MACAO systems
keep a large fraction of the object light inside the Airy disk,
stabilize the injected flux,
and a photometric estimate can be attempted. 
For commissioning purposes, a number of stars of 
various $K$-band magnitudes were observed on Nov. 3-5, 2003
with the same VLTI configuration as used for our NGC~1068 observations. 
The relation between the observed flux values and the $K$-band 
magnitudes $m_K$ 
is consistent with the expected exponential
function, so that the attempt of an absolute photometric calibration 
is reasonable.
The best-fit relations 
between $m_K$ and the photometric fluxes $P_A$ and $P_B$ (in ADU/s) are
$P_A=3.50\,10^6 \exp(-0.906\ m_K)$ and
$P_B=1.46\,10^6 \exp (-0.901\ m_K)$.
The residual dispersions on $m_K$ are $\sigma_A = 0.5$\,mag 
and $\sigma_B = 0.6$\,mag. 

Since the performance, i.e. the Strehl ratio, of the MACAO systems for
our extended source NGC\,1068 is unknown, but very likely lower than 
that of
the single bright stars (see above), the application of this  
photometric estimate on NGC\,1068 can only give us a lower limit 
of the NGC~1068 flux in our field of view (FOV) 
of $F_K \ge 130 \pm 60$\,mJy ($m_K \le 9.2 \pm 0.4$).
The size of the $K$-band Airy disk of the UTs with MACAO is 56\,mas. 
In the absence of atmospheric turbulence, the FOV
of our measurements would correspond exactly to the projection of the 
single-mode fiber on the sky. As this mode is matched by design 
to the diffraction pattern of the UTs, the effective FWHM of our FOV 
would also be 56\,mas.
In practice, however our FOV is slightly larger due to the presence of 
residual uncorrected speckles, i.e. the limited Strehl ratio.
\section{Discussion}
\begin{figure}
\centering
\resizebox{0.89\hsize}{!}{\includegraphics[angle=0]{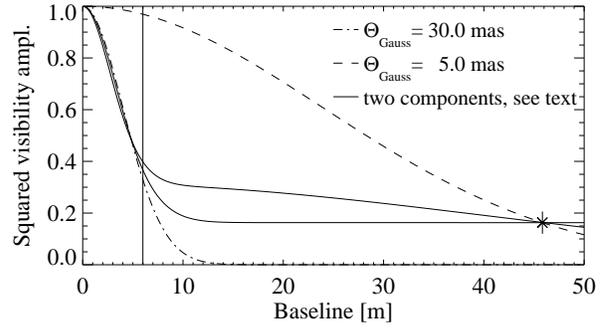}}
\caption{
Synthetic squared visibility function 
for (dashed line) a 5\,mas Gaussian, 
(dashed-dotted line) a 30\,mas Gaussian, 
and (solid lines) two examples for a two-component model 
where (upper line/ lower line) 56\%/ 40\% of the total flux comes 
from a 3\,mas/ 0.1\,mas Gaussian and the
remaining 44\%/ 60\% from a 54\,mas/ 42\,mas Gaussian (FWHM values).  
Measurements are available for baselines $B=46$\,m reported in this
letter (marked by the $\times$-symbol with error bar) 
and up to $B=6-10$\,m
(Wittkowski et al. \cite{wittkowski}; Weinberger et al. \cite{weinberger};
Weigelt et al. \cite{weigelt}).
Both measurements are only matched if the small component has a size
clearly below $\sim$\,5\,mas and if only part of the total flux in our FOV
arises from this $\lesssim$\,5\,mas component, and another part from a 
larger component.
It is assumed that the FOV is the same for all shown models.}
\label{vis_comp}
\end{figure}
\paragraph{Intensity distribution of the $K$-band emission.}
Our measured VLTI/VINCI $K$-band squared visibility amplitude 
of $|V|^2=16.3 \pm 4.3$\,\% at a projected baseline length 
of $B=45.8$\,m corresponds to a single-component Gaussian intensity 
distribution with FWHM $5.0 \pm 0.5$\,mas (taking into account 
the VINCI broad-band $K$ filter). With the distance to NGC~1068 of 14.4\,Mpc
(Bland-Hawthorn et al. \cite{bland}), this corresponds to a linear FWHM 
of $0.4 \pm 0.04$ pc. 
This means that the VLTI/VINCI visibility value is consistent with a 
multi-component model only if at least 40\% ($|V|=40.4$\,\%) of 
the flux (i.e. $\gtrsim 50$\,mJy) originates from scales $\lesssim$\,5\,mas
(Gaussian FWHM $\le$\,6.7\,mas or $\le$\,0.51\,pc at the 3 $\sigma$ level).

The $K$-band visibility values obtained by speckle interferometry
at spatial frequencies up to a baseline of 6--10\,m 
(Wittkowski et al. \cite{wittkowski}: SAO 6\,m telescope; 
Weinberger et al. \cite{weinberger}: Keck 10\,m;
Weigelt et al. \cite{weigelt}: SAO 6\,m) 
are consistent with a single-component 
Gaussian intensity distribution with azimuthally averaged FWHM
$\sim$\,30\,mas (up to $B=$\,6\,m; Wittkowski et al. \cite{wittkowski}),
and also with a larger $\sim$\,30-50\,mas component plus a much 
smaller (unresolved) component (Weinberger et al. \cite{weinberger},
Weigelt et al. \cite{weigelt}).

The comparison of the visibility measurements obtained by these
different methods is difficult since the visibility scales with the 
total observed flux in each FOV. However, since the speckle measurements 
show a structure with FWHM $\sim$\,30--50\,mas, and the VLTI/VINCI FOV
is $\sim$\,56\,mas, the total flux observed by VLTI/VINCI
is very similar to that of the compact $\sim$ 30--50\,mas speckle
component. 

Both, the VLTI/VINCI and the speckle measurements are consistent with a 
multi-component intensity distribution 
where $\gtrsim 50$\,mJy 
 originate from scales $\lesssim$\,5\,mas or $\lesssim$\,0.4\,pc (VLTI/VINCI), 
and another part of the flux from larger scales of the order 
of 40\,mas or 3\,pc (speckle). Fig. \ref{vis_comp}
shows the visibility models for a 5\,mas Gaussian matching our
VLTI/VINCI measurement, a 30\,mas Gaussian matching the speckle
measurements up to $B=$\,6\,m, and two examples for  
two-component models matching both, VLTI/VINCI and speckle measurements. 
It illustrates that two-component models are consistent with
both measurements only if the small component has a size
clearly below $\sim$\,5\,mas and a flux contribution of clearly
less than the total flux in our FOV.

In the following, we discuss the possible origin of the 
newly constrained very compact ($\lesssim$\,5\,mas) $K$-band component.
\paragraph{Sources of very compact $K$-band emission.}
Dust at the inner cavity of the torus is heated to near the 
sublimation temperature of about 1000-1500\,K and emits 
thermal $K$-band photons. These photons could escape in a 
direction other than the equatorial plane into our line-of-sight. 
Depending on the adopted dust properties and luminosity
of the central source, the sublimation radius could be as small
as about 0.3\,pc (Barvainis et al. \cite{barvainis}; 
Thatte et al. \cite{thatte}), or may as well be of the order 
of 1\,pc or above (Dopita et al. \cite{dopita}).
Hence, the size of the inner dust cavity seems to be
larger than our observed scale of FWHM clearly below $\sim$\,0.4\,pc.
It may more likely coincide with the scale of the $\sim$\,40\,\,mas
($\sim$\,3\,pc) speckle component.
However, since the work of Krolik \& Begelmann (\cite{krolikbegelmann})
it has been discussed that the dusty torus 
very likely does not have a smooth uniform shape 
but may consist of a large number of clumps (Nenkova at al. \cite{nenkova};
Vollmer et al. \cite{vollmer}). Vollmer et al. (\cite{vollmer}) 
give a theoretical estimate for the radius of such clumps 
of $r_\mathrm{Cl} \sim 0.1$\,pc. Since this size is consistent 
with our observed scale ($\lesssim$\,0.4\,pc), one might speculate 
that light from distinct clumps may contribute to our observed flux.   
Such dust clumps can also exist in polar direction at similar distances 
from the nucleus.
Another possibility is that free electrons located above and below the
broad-line emission region in the ionization cone at distances of 
only a fraction of a parsec scatter central light into our line-of-sight.  

It has already been discussed by Wittkowski et al. (\cite{wittkowski})
that a fraction of the $K$-band photons originating from
the central engine could reach us directly through only
moderate extinction in the near-infrared, despite the Seyfert 2 type
of this AGN. The $A_V$ could be as low as $\sim$\,10\,m 
(Bailey et al. \cite{bailey}), corresponding to $A_K\,\sim$1.2\,m 
assuming standard galactic extinction.
With the concept of a clumpy torus, the chance of
low extinction towards the central source is even larger than for 
a smooth dust distribution.
In this case, an analysis of the separated flux spectrum of only the 
very compact ($\lesssim$\,5\,mas) component should reveal a type 1 spectrum,
which is supported by the possible detection of a broad Br$\gamma$
line by Gratadour et al. (\cite{gratadour}).
Non-negligible $K$-band flux contributions of the order of a few hundred
milli-Jansky could arise from the central accretion flow 
(cf. Wittkowski et al. \cite{wittkowski}; 
Beckert \& Duschl \cite{beckertduschl}; Weigelt et al. \cite{weigelt}).
Stars are very unlikely to contribute significantly to the
$K$-band flux on scales $\lesssim$\,0.4\,pc 
(see e.g., Thatte et al. \cite{thatte}). 

\paragraph{Comparison to VLTI/MIDI NGC~1068 and Keck NGC~4151 observations.}
It is striking that the two NGC~1068 $K$-band scales found 
by our VLTI/VINCI observations and by speckle 
interferometry (Wittkowski et al. \cite{wittkowski}; 
Weigelt et al. \cite{weigelt}) of $\lesssim$\,0.4\,pc and 
about $2\times 4$\,pc, respectively, are very similar to those 
of the two NGC~1068 dust components reported by Jaffe et al. (\cite{jaffe}) 
based on the $N$-band VLTI/MIDI observations. The latter include 
a hot ($T>800$\,K) $0.8 \times (<1)$\,pc
and a warm ($T\sim 320$\,K) $\sim 2.5 \times 4$\,pc component.
Hot thermal emission from inner substructure of the dusty
torus, as well as direct light from the central accretion flow
could explain the $K$-band as well as the $N$-band very compact 
sub-parsec component. It is unlikely, though, that the warm 
VLTI/MIDI component and the $2\times 4$\,pc $K$-band component 
can be explained by the same dust component.

The Keck Interferometer observations of the
{\it Seyfert\,1} nucleus of NGC~4151 (Swain et al. \cite{swain}) showed
that the majority of the central ($\sim 3$\,pc) $K$-band light 
arises from very small scales of $\le 0.1$\,pc, probably from  the 
central accretion disk. Our $K$-band observations
of the {\it Seyfert\,2} galaxy NGC~1068 show that only part of the central
light arises from compact $\lesssim$\,0.4\,pc scales while another part 
arises from larger scales. This difference is in line with the unified scheme 
which predicts that the central engine of Seyfert\,2 cores
is obscured, which leads to a larger (up to 100\%) relative 
flux contribution from surrounding material. 
\section{Conclusion}
We have obtained $K$-band interferometric observations of the nucleus
of the Seyfert\,2 galaxy NGC~1068, and hereby show that the correlated
magnitude of this AGN up to a baseline of at least $\sim$\,50\,m
is such that it can be studied with the VLTI at near-infrared
wavelengths.

Together with other observations, we conclude that
a $K$-band flux of $\gtrsim$\,50\,mJy originates from scales
clearly smaller than about 5\,mas or 0.4\,pc and another part of
the flux from larger scales. Our VLTI/VINCI measurement alone
sets an upper limit of $\le$\,6.7\,mas at the 3\,$\sigma$ level
to the Gaussian FWHM of the very compact component. 
The origin of this newly constrained small-scale emission can be 
interpreted as substructure of the dusty torus, as for instance 
part of a clumpy inner cavity or distinct clumps forming the torus, 
or as direct emission from the central accretion flow 
viewed through only moderate extinction in the $K$-band.
\begin{acknowledgements}
The VLTI observations reported here were made possible through the 
efforts of the whole ESO VLTI and MACAO commissioning teams.
%the latter including R. Donaldson and E. Fedrigo, M. Kasper.
\end{acknowledgements}

\end{document}